\documentclass[twocolumn]{aastex631}

\usepackage{amsmath}
\usepackage{subfigure}
\definecolor{blue-violet}{rgb}{0.30, 0.1, 0.89}

\begin{document}

\title{Polarization of Intrabinary Shock Emission in Spider Pulsars}

\author[0000-0002-0786-7307]{Andrew G. Sullivan}
\affiliation{Kavli Institute for Particle Astrophysics and Cosmology, Department of Physics, Stanford University, Stanford, CA 94305, USA}

\author[0000-0001-6711-3286]{Roger W. Romani}
\affiliation{Kavli Institute for Particle Astrophysics and Cosmology, Department of Physics,
Stanford University, Stanford, CA 94305, USA}



\begin{abstract}
In `spider' pulsars, the X-ray band is dominated by Intrabinary Shock (IBS) synchrotron emission. While the double-peaked X-ray light curves from these shocks have been well characterized in several spider systems (both black widows and redbacks), the polarization of this emission is yet to be studied. Motivated by the new polarization capability of the Imaging X-ray Polarization Explorer (IXPE) and the confirmation of highly ordered magnetic fields in pulsar wind nebulae, we model the IBS polarization, employing two potential magnetic field configurations: toroidal magnetic fields imposed by the pre-shock pulsar wind, and tangential shock-generated fields, which follow the post-shock flow. We find that if IBSs host ordered magnetic fields, the synchrotron X-rays from spider binaries can display a high degree of polarization ($\gtrsim50\%$), while the polarization angle variation provides a good probe of the binary geometry and the magnetic field structure. Our results encourage polarization observational studies of spider pulsars, which can distinguish the proposed magnetic models and better constrain unique properties of these systems.
\end{abstract}


\section{Introduction} \label{sec:intro}

Spider binaries contain a millisecond pulsar and a low-mass companion star in a tight orbit with period $P_b\lesssim1$ day.  Spiders are often classed as black widows \citep{1988Natur.333..237F}, with sub-stellar $M_c<< 0.1$ M$_\odot$ companions or as redbacks \citep{2013IAUS..291..127R}, with $M_c\approx 0.1-0.4$ M$_\odot$. In these systems, the pulsar irradiates the companion and  drives a stellar wind \citep{1988Natur.334..225K,1988Natur.334..684V}.
The relativistic pulsar wind and massive companion wind collide, forming an intrabinary shock (IBS). These sources emit across the electromagnetic spectrum, with radio and gamma-ray emission from the pulsar itself, optical emission from the companion star and X-ray emission dominated by the IBS. In redbacks, the companion wind momentum dominates that of the pulsar, causing the IBS to wrap around the pulsar, while in black widows, the IBS wraps around the companion \citep{2016ApJ...828....7R}.

The pulsar wind is strongly magnetized, so the shock-accelerated particles emit prominent synchrotron X-rays in the post-shock flow \citep{1993ApJ...403..249A, 2019ApJ...879...73K, 2020ApJ...904...91V, 2021ApJ...917L..13K}.
This flow accelerates to mildly relativistic speeds, so that the non-thermal IBS orbital light curves often display two caustic peaks, associated with Doppler-beamed emission from relativistic particles traveling tangent to the instantaneous line of sight. The light curve is quite sensitive to the geometry of the IBS, with the wind momentum ratio and observer viewing angle determining the separation of the two peaks. 
Conversely, the IBS spectrum reveals much about the acceleration and cooling of the particles in the post-shock flow. 

Synchrotron emission from organized fields is polarized. While synchrotron X-ray light curve and spectral analyses have been conducted on a number of redbacks and black widows \citep[e.g.][]{2014ApJ...783...69G, 2016ApJ...828....7R, 2019ApJ...879...73K, 2021ApJ...917L..13K, 2023ApJ...952..150P}, the X-ray polarization of these sources has not yet been studied. New X-ray capabilities such as IXPE  \citep{2016SPIE.9905E..17W, 2022JATIS...8b6002W} may make this possible, affording new information on the IBS magnetic field geometry. IXPE has already demonstrated that pulsar wind nebulae (PWNe) have very high synchrotron polarization \citep{2022Natur.612..658X, 2023NatAs...7..602B}, suggesting similar features in spiders, which have even harder spectra. 

In this paper, we present a pilot study of IBS polarization, exploring some simple field models (Sec. \ref{sec:PolarizationModel}), and computing the expected polarization signatures in realistic spider geometries (sec.\,\ref{sec:Results}). {Our models use a {semi}-analytic thin shock treatment, with analytic expressions for the IBS contact discontinuity shape as well as particle radiation and cooling, as in \citet{2019ApJ...879...73K}. This allows for rapid model generation to explore parameter space and fit data. These models capture well the observed IBS pulse shapes and spectra and fits can constrain a system's geometrical parameters, although omitting the detailed post-shock spreading that would be captured in magnetohydrodynamic (MHD) models}. The prospects for detection and the potential for deeper probes of the IBS structure are briefly discussed in Sec.\,\ref{sec:conclusion}. 

\section{IBS Synchrotron Polarization}
\label{sec:PolarizationModel}

In the IBS, shocked pulsar wind electrons and positrons become accelerated to very high energies \citep{2011ApJ...741...39S,2022ApJ...933..140C} and flow along the shock surface \citep{2008MNRAS.387...63B, 2015A&A...577A..89B, 2023PhRvE.107b5201M}. The magnetic fields, either remnants of the shock-compressed, incompletely cancelled striped wind fields, or MHD instability-generated fields, stretched along the post-shock flow, allow particle cooling via synchrotron radiation. If sufficiently uniform, this field induces high polarization \citep{1959ApJ...130..241W, 1979rpa..book.....R} perpendicular to the projection of the magnetic field on the sky. While the complex 3-D structure of the IBS means that such field directions vary, the beaming of the caustic peaks guarantees that a sub-set of these directions dominates at a given orbital phase. This raises the prospect of substantial net polarization and motivates a detailed computation of the IBS synchrotron emission. 

\subsection{Polarized Synchrotron Radiation}

The total synchrotron power spectrum {per particle} is \citep[e.g.]{1979rpa..book.....R, 2019ApJ...879...73K}
\begin{equation}
\label{eq:syncpower}
   P(\omega, \gamma)=\frac{\sqrt{3} q^3 B \sin\alpha}{2\pi mc^2} F\left(\frac{\omega}{\omega_c(\gamma)}\right),
\end{equation}
where $q$ and $m$ are the charge and mass of the radiating particle, $B$ is the magnetic field strength, $\alpha$ is the pitch angle between the particle velocity and the magnetic field, $\omega_c(\gamma)\equiv3qB \gamma^2 \sin\alpha /2mc$ is the characteristic synchrotron frequency, $\gamma$ is the particle Lorentz factor, $F(x)\equiv x\int_x^\infty K_{5/3}(y)dy$, and $K_{n}(x)$ is the modified Bessel function of order $n$. The power radiated along and perpendicular to the magnetic field direction are
\begin{subequations}
    \begin{equation}
           P_{||}(\omega, \gamma)=\frac{\sqrt{3} q^3 B \sin\alpha}{4\pi mc^2} \left[F\left(\frac{\omega}{\omega_c}\right)-G\left(\frac{\omega}{\omega_c(\gamma)}\right)\right],
    \end{equation}
   \begin{equation}
           P_{\perp}(\omega, \gamma)=\frac{\sqrt{3} q^3 B \sin\alpha}{4\pi mc^2} \left[F\left(\frac{\omega}{\omega_c}\right)+G\left(\frac{\omega}{\omega_c(\gamma)}\right)\right],
    \end{equation}
\end{subequations}
where $G(x)\equiv x K_{2/3}(x)$. The emission projected to the sky direction $\vec{n}$ will be polarized with polarization vector $\vec{e}=\vec{n}\times \vec{b}$, where $\vec{b}$ is the magnetic field direction vector. For an angle $\chi$ between the magnetic field and a particular reference direction in the plane of the sky, the Stokes parameters for linear polarization {from an individual emission zone} \citep[e.g.,]{2014JKAS...47...15T} are
\begin{subequations}
    \begin{equation}
          Q=I(0)-I\left(\frac{\pi}{2}\right)=(P_\perp-P_{||})(\sin^2\chi-\cos^2\chi),
    \end{equation}
   \begin{equation}
   \begin{split}
           U=&I\left(\frac{\pi}{4}\right)-I\left(\frac{3\pi}{4}\right)\\&=
           (P_\perp-P_{||})\left[\sin^2\left(\chi-\frac{\pi}{4}\right)-\cos^2\left(\chi-\frac{\pi}{4}\right)\right],
    \end{split}
    \end{equation}
    \label{eq:singlezone}
\end{subequations}
where $I(\theta)$ is the intensity along the direction with angle $\theta$ to the reference direction. The polarization degree $\Pi$ in this case is \citep[e.g.]{1979rpa..book.....R}
\begin{equation}
    \Pi =\frac{\sqrt{Q^2+U^2}}{I_{tot}}=\frac{P_{\perp}(\omega)-P_{||}(\omega)}{P_{\perp}(\omega)+P_{||}(\omega)},
\end{equation}
where $I_{tot}$ is the total intensity.

When the bulk velocity of the emitting region is nonzero, the emission will be relativistically boosted. The boosting also affects the polarization direction. For an emitting particle population traveling in direction $\hat{v}$ with bulk Lorentz factor $\Gamma$, the radiated power will be boosted by 
\begin{equation}
    P_{obs}(\omega)=D^{-3} P\left(D \omega\right),
\end{equation}
with $D\equiv\Gamma \left[1-(1-\Gamma^{-2})^{1/2}\hat{v}\cdot \vec{n}\right]$. $P_\perp$ and $P_{||}$ are boosted in the same manner. The polarization vector in the observer frame $\vec{e}_{obs}$ is transformed to \citep{1979ApJ...232...34B, 2003ApJ...597..998L, 2018ApJ...864..140P}
\begin{equation}   
\label{eq:polvector}\vec{e}_{obs}=\frac{\vec{n}\times\vec{q}}{\sqrt{\vec{q}^2-(\vec{n} \cdot \vec{q})^2}}, 
\end{equation}
where $\vec{q}\equiv\vec{b}+\vec{n}\times[(1-\Gamma^{-2})^{1/2}\hat{v}\times \vec{b}]$ for $\vec{b}$ defined in the observer frame.

\subsection{Intrabinary Shock Model}
We adopt the IBS model of \cite{2019ApJ...879...73K} and add polarization as outlined above. {This model is semi-analytic and designed to capture the impact of geometry on the resulting emission.} The primary geometry of the IBS is governed by the stellar wind to pulsar wind momentum ratio $\beta=\dot{M_w}v_w c/\dot{E}_{PSR}$ \citep{2016ApJ...828....7R, 2019ApJ...879...73K}. The $\beta>1$ case generally corresponds to redbacks, while the $\beta<1$ case to black widows. The exact shape of the contact discontinuity between the shocks may also depend on the latitudinal distribution of the pulsar wind \citep{ 2019ApJ...879...73K}. The simplest case occurs when the pulsar wind is spherical, and the geometry of the IBS is given by \cite{1996ApJ...469..729C}; if the pulsar wind is equatorial, the appropriate formulae are given by \cite{2019ApJ...879...73K}. The shape of the IBS is also distorted by sweepback due to the companion's orbital motion. This effect is parameterized by $f_v=v_w/v_{orb}$, where $v_{orb}$ is the orbital speed \citep{2016ApJ...828....7R}. For small $f_v$, the shock contact discontinuity will trace out an Archimedean spiral \citep{2008MNRAS.388.1047P, 2012A&A...546A..60L, 2015A&A...577A..89B}.

We extend the ICARUS IBS code \citep{2012ApJ...748..115B, 2016ApJ...828....7R, 2019ApJ...879...73K} to include synchrotron polarization. Computationally, the IBS {is assumed to be a thin shock along the} contact discontinuity, which is divided into triangular tiles {of constant angular size as viewed from the pulsar, }representing different zones from which synchrotron radiation is emitted. {In this paper, we assume a spherical pulsar wind for simplicity.} At the IBS, the pulsar wind injects an electron and positron population with energy spectrum {in the flow frame}
\begin{equation}
\label{eq:PL}
    N(\gamma_e) d\gamma_e=N_0 \gamma_e^{-p} d\gamma_e,
\end{equation}
where $\gamma_e$ is the electron/positron Lorentz factor in range $\gamma_{min}<\gamma_e<\gamma_{max}$, $N_0$ (in e/${\rm cm^2/s}$) is a global normalization coefficient and $p$ depends on the particle acceleration mechanism. {$N_0$ is typically a free parameter in IBS fits to data; the relative normalization of each tile is $N_{0, j}\propto1/r_j^2$, where $r_j$ is the distance between the $j$th tile and the pulsar. When the corresponding energy flux of the power law is integrated over solid angle, it can be usefully compared with the pulsar spin-down power ${\dot E} = I\Omega {\dot \Omega}$.} After injection, the particles duct from an individual tile downstream and radiatively cool. The bulk velocity of the electron/positron population is approximated with direction $\hat{v}$ tangent to the contact discontinuity. Bulk Lorentz factor increases along the shock, approximated as
\begin{equation}
\Gamma_B(s)=\Gamma_0\left(1+k\frac{s}{r_0}\right),
\end{equation}
where $s$ is the arclength from the nose to a given tile, $r_0$ is the nose-standoff distance from the pulsar, $\Gamma_0$ is the Lorentz factor at the nose, and $k$ is a scaling parameter that controls the flow velocity increase. The magnetic field $\vec{B}(r)$ is defined at each tile of the IBS as a function of distance from the pulsar {and boosted to the flow frame at each tile when computing the emission.} We defer further discussion of the magnetic field configuration to sec. \ref{subsec:magneticfield}.

{We compute the residence time of the bulk flow in each tile in the flow frame
\begin{equation}
    \tau_j=\frac{d_j}{c\Gamma_j \sqrt{1-\Gamma_j^{-2}}},
\end{equation}
where $d_j$ is the physical length of the tile. Synchrotron cooling evolves the flow frame particle spectrum as \citep{1979rpa..book.....R, 2019ApJ...879...73K}
\begin{equation}
\label{eq:synchcool}
 \gamma_e(t)=\gamma_{e,0}\left(1+\frac{2e^4 B^2\sin^2\alpha}{3 m^3 c^5}\gamma_{e, 0}t\right).
\end{equation} 
We assume the time spent by the electrons/positrons in a particular tile is evenly distributed over $(0, \tau_j)$. Since the tiles are triangular, we add half of the final cooled particle spectrum after $\tau_j$ to the freshly injected electron spectrum of each of the two downstream `daughter' tiles. In our thin shock approximation, we do not include possible adiabatic losses.} 

The synchrotron spectrum projected in a given sky direction from tile $j$ is computed as 
\begin{equation}
L_j(\omega)=D_j^{-3}\int_{\gamma_{min}}^{\gamma_{max}} N_j(\gamma_e) P(D_j\omega, \gamma_e)d\gamma_e,
\end{equation}
with $P(\omega, \gamma_e)$ from Eq.\,\ref{eq:syncpower} and $D$ defined above. The {injected} $e^\pm$ power law (Eq.\,\ref{eq:PL}) gives rise to a photon spectrum ${\rm d}N_\gamma/{\rm d}E \sim E^{-\Gamma_x}\sim E^{-(p-1)/2}${; more generally $N_j(\gamma_e)$ for a tile is the full electron population, including cooled electrons advected from upstream tiles}. For the emission polarized perpendicular and parallel to the projected magnetic field on the sky in a given sky direction $\vec{n}$, one replaces $P$ in Eq.\,\ref{eq:syncpower} with $P_\perp$ and $P_{||}$ of Eq.\,2.
The total power on the sky in a given band emitted by tile $j$ is simply
   \begin{equation}  L_j=\int_{\omega_{min}}^{\omega_{max}}L_j(\omega)d\omega,
\end{equation}
where $\omega_{min}$ and $\omega_{max}$ are the minimum and maximum frequency in the spectral band of interest.

The observed polarization vector from each tile is given by Eq.\,\ref{eq:polvector}. We set $\hat{l}_{proj}$, the orbital angular momentum projected on the sky, as our reference direction. The Stokes $Q$ and $U$ from each tile are then
\begin{subequations}
\begin{equation}
\begin{split}
    Q_j=&\left(L_{ \perp, j} -L_{||, j}\right)\\&\times\left[(\vec{e}_{obs, j} \cdot \hat{l}_{proj})^2-(\vec{e}_{obs, j} \cdot[\vec{n} \times\hat{l}_{proj}])^2\right],  
\end{split}
\end{equation}
\begin{equation}
\begin{split}
    U_j=&\left(L_{ \perp, j} -L_{||, j}\right)\\&\times\left[(\vec{e}_{obs, j} \cdot \hat{l}_{proj, 45^\circ})^2-(\vec{e}_{obs, j} \cdot[\vec{n} \times\hat{l}_{proj, 45^\circ}])^2\right], 
\end{split}
\end{equation}
\end{subequations}
 where $\hat{l}_{proj, 45^\circ}$ is the unit vector rotated counter clockwise $45^\circ$ from 
 $\hat{l}_{proj}$. The total luminosity and Stokes parameters of the IBS are
 \begin{subequations}
 \label{eq:flux}
   \begin{equation}
    L=\sum_j L_j,
\end{equation}
    \begin{equation}
    Q=\sum_j Q_j,
\end{equation}
    \begin{equation}
    U=\sum_j U_j.
\end{equation}
\end{subequations}
The polarization degree and angle on the sky are then given by
\begin{subequations}
  \begin{equation}
    \Pi_{tot}= \frac{\sqrt{Q^2+U^2}}{L},
\end{equation}
  \begin{equation}
    \chi_{tot}= \frac{1}{2}\arctan_2{\left(\frac{U}{Q}\right)}.
\end{equation}  
\end{subequations}
{Note that $\chi_{tot}$ here is the polarization angle of the summed emission, while $\chi$ in eq. \ref{eq:singlezone} is the single zone polarization angle.}
Evaluating Eqs.\,\ref{eq:flux} for all $\vec{n}$ gives skymaps \cite[see Fig.\,3 of][]{2019ApJ...879...73K} of flux, $Q$, and $U$. Selecting a particular inclination angle $i$ between $\vec{n}$ and the binary orbital angular momentum gives the phase-resolved light curve, polarization degree, and polarization angle. {Light curves have orbital phases $0\leq\Phi<1$, with $\Phi=0$ the pulsar ascending node} {(pulsar inferior conjunction at $\Phi=0.75$ has the pulsar between us and the companion).}

\subsection{Magnetic field geometry}
\label{subsec:magneticfield}
Outside the light cylinder, the pulsar wind magnetic field should be toroidal (i.e. $B(r)\propto r^{-1}$) and takes the form of an MHD wind with stripes of alternating polarity separated by current sheets \citep{1990ApJ...349..538C, 2015MNRAS.448..606C}. At the small IBS distance, we expect stronger incident fields than those of PWNe termination shocks \citep[e.g.][]{2012SSRv..166..231R}. At the shock, magnetic reconnection likely occurs, accelerating shocked particles; {a residual magnetic field remains in the IBS after the stripes annihilate} \citep{2003MNRAS.345..153L, 2007A&A...473..683P, 2011ApJ...741...39S}.
The very hard spectra observed in spiders (i.e.\,$\Gamma_x\approx 1$) \citep{2014ApJ...781....6B, 2021ApJ...917L..13K, 2023ApJ...952..150P} supports this picture \citep{2022ApJ...933..140C, 2023arXiv230212269Z}. The toroidal field remaining after pulsar wind annihilation represents a promising candidate for the magnetic field geometry, but post-shock dynamics such as turbulence may produce differently ordered fields \citep[e.g.][]{2016MNRAS.461..578G}. 
\begin{figure}
   \includegraphics[width=\columnwidth]{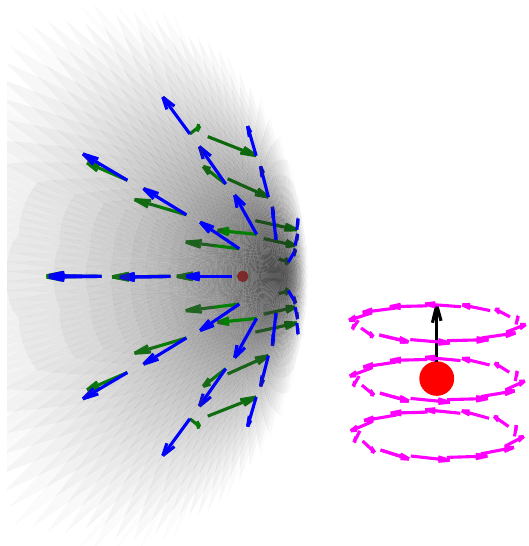}
    \caption{A 3-D visualization of the IBS flow and magnetic structure we consider in this paper. {Here the pulsar (red dot), in this case a redback, is wrapped by the IBS (grey fading surface)}. The IBS is viewed from inclination $i=90^\circ$ at orbital phase $\Phi=0.58$. Blue arrows mark the bulk flow directions $\hat{v}$ and the green arrows show the magnetic field vectors $\vec{b}$ for the shock-modified cylindrical model. For the flow model, the blue arrows also represent the magnetic field vectors. Green arrows with heads ending on the blue vectors are from the far side, viewed through the IBS. {The lower right shows the pre-shock cylindrical field around the pulsar. The magenta rings denote the magnetic field lines while the black arrow represents the orbital angular momentum axis; these strongly recycled spider pulsars are assumed to be spin-aligned.}}
    \label{fig:MagneticVisual}
\end{figure}

We assume here that the field magnitude immediately pre-shock is 
\begin{equation}
\label{eq:magneticfield}
    B(r)= B_0\left(\frac{r_0}{r}\right),
\end{equation} 
where $B_0$ is the value of the magnetic field at the IBS nose, $r_0$ is the distance from the pulsar to the nose, and $r$ is the distance from the pulsar to a point on the IBS. For typical inferred $B_0$, synchrotron losses can be significant for energetic particles radiating at high energies {as they flow along the shock}. There can also be spectral features associated with the electron spectrum cutoff at $\gamma_{\rm max}$. These features are often in the hard X-ray/soft $\gamma$-ray band for typical spider parameters \citep{2019ApJ...879...73K}.

The magnetic field structure inside an IBS is not yet understood. To illustrate the range of uncertain IBS magnetic field structures, we present two simple model geometries: 1) a shock-modified cylindrical field and 2) a flow geometry. The cylindrical magnetic field geometry is inspired by the toroidal structure already observed in PWNe \citep{2022Natur.612..658X, 2023NatAs...7..602B} and corresponds to the toroidal field imprinted on the shock by the partially annihilated striped wind \citep{1990ApJ...349..538C, 1999A&A...349.1017B, 2007A&A...473..683P}. In this model, we set $\vec{b}=\hat{\phi}$ in the pre-shock wind with $\hat{\phi}$ azimuthal in a cylindrical system centered on the pulsar {with cylindrical axis along the orbital angular momentum (see the inset in Fig.\,\ref{fig:MagneticVisual})}. Note that the IBS shock, unlike a PWN shock, is oblique in many regions. In the immediate post-shock IBS, the component of $\vec{B}$ parallel to the shock surface is magnified by a factor of $3$ (for an ultra-relativistic shock with adiabatic index $\gamma=4/3$), while the normal component is unaltered. This drives ${\vec b}$ closer to parallel in the shocked wind. We assume here that the radiation arises in this immediate post-shock zone. 

Alternatively, in the flow model, we envision a scenario in which field lines become stretched along the motion of particles flowing in the shock. This is motivated by the radially stretched magnetic field structure inferred in young supernova remnants \citep{1976AuJPh..29..435D, 2015A&ARv..23....3D, 2022ApJ...938...40V} {as well as the field advection seen in IBS particle-in-cell simulations \citep[e.g.][]{2022ApJ...933..140C}}. Geometrically, we assume that the magnetic field follows the IBS flow so that $\vec{b}=\hat{v}$, while the magnitude is the pre-shock value given by eq. \ref{eq:magneticfield} multiplied by 3 to account for shock compression. We show a 3-D visualization of the IBS and the magnetic field configurations in Fig. \ref{fig:MagneticVisual}. {Note that these geometries are specified in the lab frame. The fields are boosted to the flow frame when evaluating eq. \ref{eq:syncpower}.} These two magnetic field geometries have maximal projected angle differences toward the orbital poles. While phenomenological, they illustrate a wide range of possible polarization behavior.

\section{Model Results}
\label{sec:Results}
We will illustrate the IBS peak and polarization pattern with a `redback' geometry. In this case the IBS lies close to the pulsar and the wind compresses and accelerates the flow, leading to more prominent IBS peaks. Redback sources generally also have larger X-ray fluxes \citep[e.g.][]{2023MNRAS.525.3963K}, so they will be prime targets for IBS polarization studies. Black widows should have broadly similar polarization behavior, except with peak features centered on orbital phase $\Phi=0.25$ rather than $\Phi=0.75$. The larger obliquity across much of the black widow IBS will cause some differences for the cylindrical model.

\begin{figure*}
    \includegraphics[width=\linewidth]{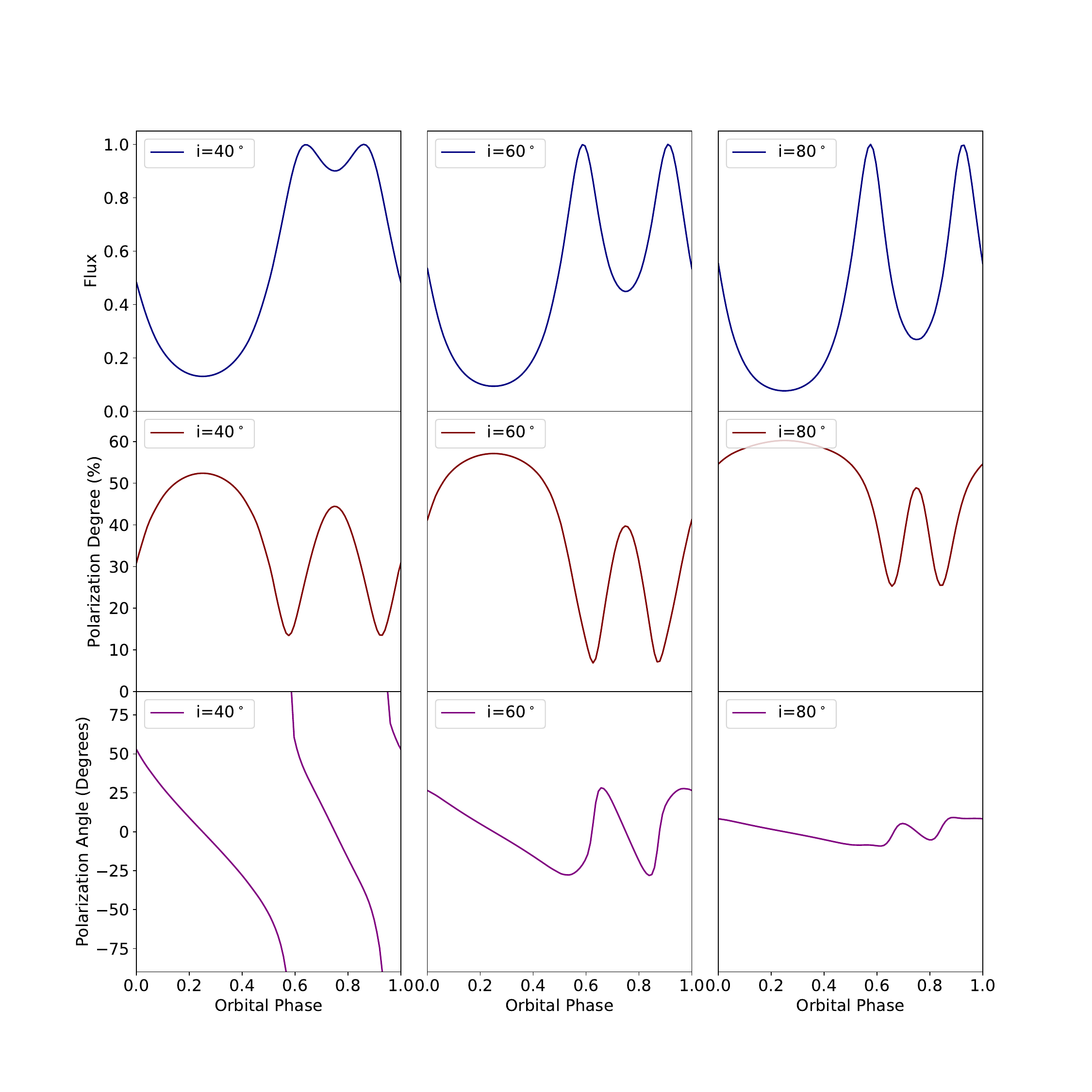}
    \caption{Plots of the IBS flux (top), polarization degree (middle), and polarization angle (bottom) for $i=40^\circ$ (left), $i=60^\circ$ (center), and $i=80^\circ$ (right) for the cylindrical magnetic field model in the uncooled power law part of the spectrum. Polarization angle $0^\circ$ corresponds to the polarization vector $\vec{e}$ aligned with the binary orbital angular momentum vector projected onto the sky plane. The IBS model was computed with $\beta=5$, $f_v=\infty$, $p=1$, $\Gamma_0=1.1$, and $k=0.2$.}
    \label{fig:Cylindrical}
\end{figure*}

\begin{figure*}
    \centering
    \includegraphics[width=\linewidth]{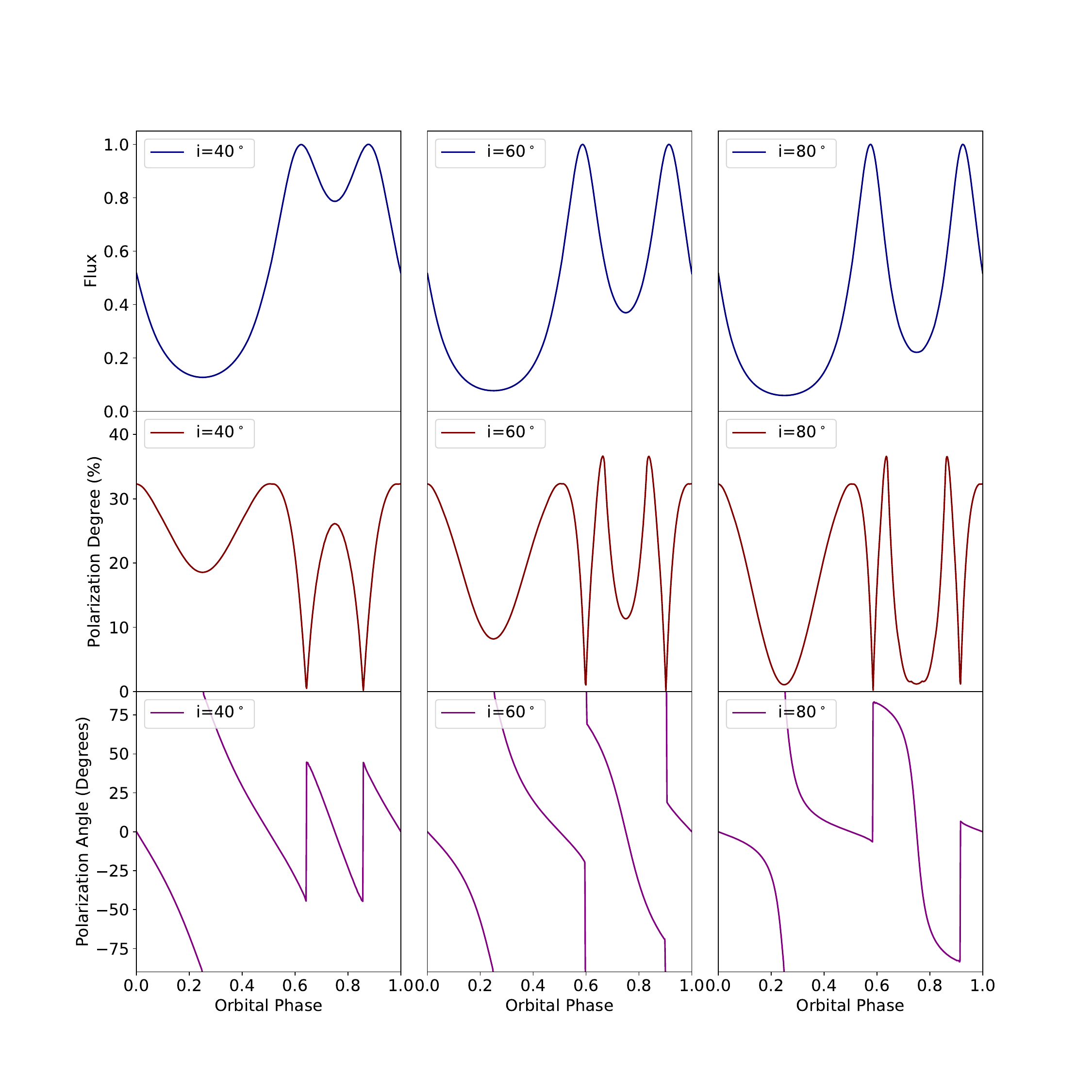} 
    \caption{Plots of the IBS flux (top), polarization degree (middle), and polarization angle (bottom) for $i=40^\circ$ (left), $i=60^\circ$ (center), and $i=80^\circ$ (right) for the flow magnetic field model. The conventions and model parameters are the same as those in Fig. \ref{fig:Cylindrical}}
    \label{fig:Flow}
\end{figure*}

In figs.\,\ref{fig:Cylindrical} and \ref{fig:Flow}, we show the predicted polarization properties with the cylindrical and flow magnetic field models for a range of inclination angles. In both cases, the {model parameters are} chosen so that the emission is from the uncooled power law spectral component, {as typically appropriate for the soft X-ray band \citep{2019ApJ...879...73K}}. The polarization profiles of these two models are notably different. Most prominently, the polarization degree is substantially higher in the cylindrical model. The polarization degree is highest around the flux minimum but can be substantial between the caustic peaks. The total polarization degree increases with $i$, as the viewing angle more closely coincides with the equatorial plane in which most field lines lie. {The orbital variation in the electric vector polarization angle (EVPA) decreases with $i$, as polarization direction aligns more closely with the orbital angular momentum.} The fastest EVPA sweep and minimum polarization degree occur across the two caustic peaks. Note that the polarization degree minima lie outside the flux peaks for small $i$, inside for large $i$.


\begin{figure*}
    \centering
    \includegraphics[width=\linewidth]{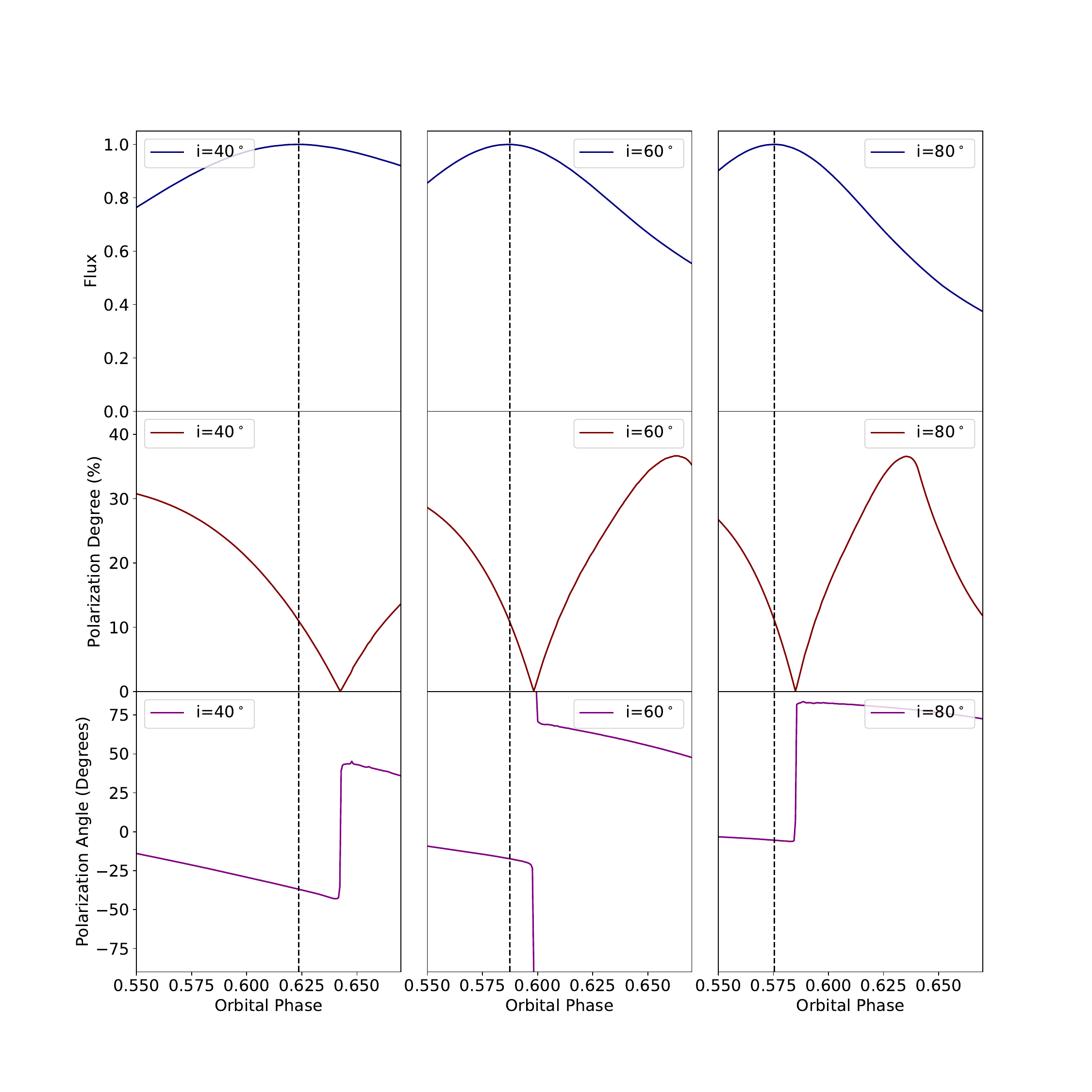}  \caption{Plots of the flow magnetic field model flux, polarization degree, and polarization angle as in Fig. \ref{fig:Flow}, zoomed in around the first peak. The dashed line in each subplot marks the orbital phase of the flux peak.}
    \label{fig:FlowZoom}
\end{figure*}

The orbital peak emission comes from IBS zones with flow lines near-tangent to the line of sight, due to relativistic beaming. For the flow model, the position angle of the projection of the local magnetic field on the sky varies rapidly about this tangent point, both on the IBS surface and with binary phase. This causes the rapid EVPA sweep near the peaks, and strong depolarization of the integrated emission. Fig. \ref{fig:FlowZoom} highlights the phase region surrounding the first peak in the flow model, showing the polarization behavior. Notice that the polarization degree minimum is slightly behind the first peak phase (and slightly ahead of the second peak). Thus the phase where the polarized flux most nearly cancels is offset from the flux peaks in this model as well, although the shifts are more subtle. In detail, the offset is sensitive to the IBS surface curvature, field structure, and emissivity in the zones where the caustic peaks form. Since Doppler boosting at bulk Lorentz factor $\Gamma_B$ mixes caustic emission over a beaming angle $1/\Gamma_B$, the maximum phase offset scales with $P_b/\pi\Gamma_B$. 

\begin{figure*}
    \centering
    \includegraphics[width=\linewidth]{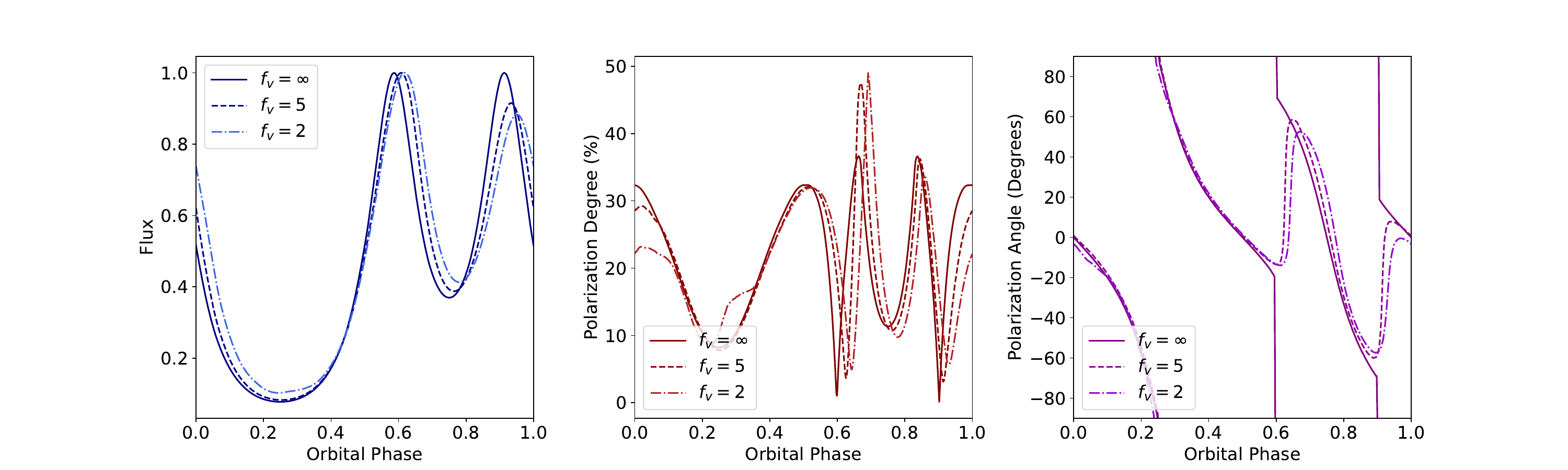}
    \caption{Sweepback-induced distortion of IBS light curves (left), polarization degree (middle), and polarization angle (right) {with the same parameters (except  $f_v$) as Fig. \ref{fig:Flow}}. $f_v$ describes the companion's wind speed and, hence, is inversely proportional to the sweepback distortion. Curves show the flow magnetic field model at $i=60^\circ$.}
    \label{fig:fv5}
\end{figure*}

While the detailed profiles depend on the peculiarities of the IBS structure and particle flow, a few basic physical effects can distort the peaks in interesting ways. Orbital motion can impact the light curves and polarization profiles for low companion wind speeds. Lower values of $f_v$ generate asymmetry in the height of the two light curve caustic peaks as the IBS is swept back in an Archimedean spiral \citep[e.g.]{2016ApJ...828....7R, 2019ApJ...879...73K}. The phase-resolved polarization profile exhibits similar features with higher polarization around the higher peaked caustics. We show an example light curve and phase-resolved polarization profile for the flow magnetic field model with different values of $f_v$ and $i=60^\circ$ in Fig. \ref{fig:fv5}. For the cylindrical model the polarization changes little from the $f_v=\infty$ case.

\begin{figure*}
    \centering
    \includegraphics[width=\linewidth]{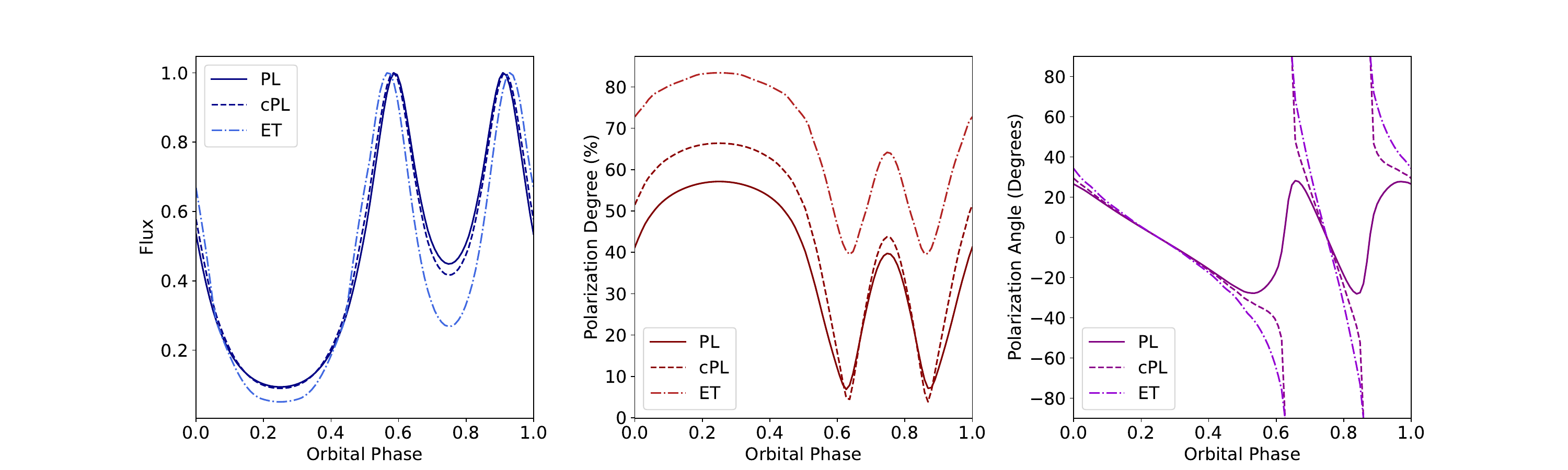}
    \caption{Effects of synchrotron spectral features on the light curve (left), polarization degree (middle), and polarization angle (right), here for the cylindrical model at $i=60^\circ$ {with the same parameters as Fig. \ref{fig:Cylindrical}}. The line types show spectral bands dominated by three spectral regimes: the low energy uncooled power law (PL), the intermediate range where a cooled power law (cPL) dominates, and a high energy range sensitive to the exponential tail (ET) of the IBS spectrum.  Note that the polarization degree grows as the spectrum steepens at high energy.}
    \label{fig:spectra}
\end{figure*}

The IBS spectrum, peak shape, and polarization are also affected by synchrotron cooling and the upper cutoff $\gamma_{\rm max}$. While the uncooled power law population dominates the spectrum at low energies, the spectrum has a cooling break after the particles flow a distance $s$ with $\Gamma_B\sim1.2$ at $E\simeq 10 \left(\frac{B}{10 \text{ G}}\right)^{-3}\left(\frac{s}{5\text{ R}_\odot}\right)^{-2}$\,keV. In addition, the spectrum cuts off exponentially above $E\simeq170\left(\frac{\gamma_{max}}{10^6}\right)^2\left(\frac{B}{10 \text{ G}}\right)$\,keV due to the maximum particle energy.  In Fig. \ref{fig:spectra}, we compare light curves and polarization profiles for the three spectral regimes for one IBS geometry. The peak separation widens while the peaks themselves narrow with energy \citep{2019ApJ...879...73K}. The polarization structure remains similar across the three regimes, but, interestingly, the polarization degree increases at the higher energies. In the cooled power law regime, the polarization degree increases from $50\%$ to 65\%, while in the exponential tail, the polarization degree reaches 80\%. This arises from $F(\omega/\omega_c)$ and $G(\omega/\omega_c)$ for $\omega>\omega_c$.  The EVPA sweep loses the reversal at energies above the cooling break.

\section{Discussion and Conclusion}
\label{sec:conclusion}
The high polarization we predict for the IBS synchrotron emission makes spider pulsars interesting polarization targets. As the IBS emission dominates X-rays, IXPE may be able to probe the polarization for X-ray bright spiders. {The polarization levels predicted here might be considered upper limits, since strong turbulence will decrease the observed polarization degree. Since IXPE has shown that PWN termination shocks have polarizations approaching the maximum turbulence-free levels allowed for synchrotron emission, however, the same may apply here. Conversely the IBS components of spider pulsars, while dominant in the X-rays, are faint compared to typical accretion powered X-ray sources; only a few spiders reach the $f_{2-8{\rm keV}} \approx 10^{-12} {\rm\text{ } erg/cm^2/s}$ fluxes required for high-significance IXPE measurements in plausible Ms exposure times.} 

In IXPE's 2-8 keV bandwidth, we should typically be mapping uncooled synchrotron emission, although some spider parameters imply $B_0 > 100$\,G, moving the cooling break into the soft X-ray band. Future Compton polarimeters, sensitive to hard X-rays and low energy gamma-rays \citep{1997SSRv...82..309L} may also probe the cooled population and possibly the exponential tail, especially given the increasingly high polarization expected. NuSTAR hard X-ray observations of some redbacks have shown double peak structure consistent with the IBS \citep[e.g.][]{2018MNRAS.478.3987K, 2023ApJ...952..150P}, making targeted polarization measurements of the IBS in this band appealing.

The uncooled power law spectrum of the IBS can extend down to the optical band. In such cases, a non-thermal flux component, most prominent at bluer wavelengths, adds to the companion thermal emission. Such non-thermal fluxes have already been observed in some spiders \citep[e.g.][]{2020ApJ...902L..46N}. As the companion thermal emission is unpolarized, the polarized IBS contribution will generally be strongly diluted, but may be accessible with high precision optical polarimetry.

{Our calculations may also be relevant to other binaries showing hard-spectrum IBS emission, such as high mass $\gamma$-ray binaries containing energetic pulsars. Unlike spiders, where our light curve and polarization computations are for orbital phase-varying views of a stationary IBS, these systems are typically eccentric, with the IBS structure modulated by changing orbital separation and/or equatorial disk crossings. This complicates the orbital light curve and polarization modulation; however, if the post-shock magnetic field takes on the organized structures assumed in this paper, some phases may also show high polarization.}

Our models do not exhaust the possible field structures and other effects may have a significant impact on polarization behavior. For example, field annihilation in the pulsar wind may be a strong function of pulsar co-latitude and the residual field strength may be sensitive to pulsar spin-orbit misalignment \citep{1999A&A...349.1017B}. {Additionally, turbulence and disordered magnetic fields can naturally decrease the overall polarization. More realistic magnetic field models which account for these behaviors will require detailed MHD simulations.} We defer discussion of more detailed field configurations to future work. {The models presented here do, however, illustrate possible phase-resolved polarization degree and EVPA features.} In conclusion, the highly ordered fields seen in PWNe suggest the same may be true for IBSs; the resulting high polarization can provide a powerful diagnostic of magnetic reconnection and particle acceleration in these systems.

\section*{Acknowledgements}
The authors are grateful to Roger Blandford for useful discussions {and the anonymous referee for a prompt, useful review}. This work was supported in parts by NASA grants NMM17AA26C and 80NSS229K1506. A.S. acknowledges the support of the Stanford University Physics Department Fellowship and the National Science Foundation Graduate Research Fellowship Program.
\bibliography{refs}{}
\bibliographystyle{aasjournal}

\end{document}